\documentclass[conference]{IEEEtran}
\IEEEoverridecommandlockouts
\usepackage{aaai20}
\usepackage{amsmath,amssymb,amsfonts}
\usepackage{algorithmic}
\usepackage{graphicx}
\usepackage{textcomp}
\usepackage{xcolor}
\usepackage{autobreak}
\usepackage{svg}
\usepackage{amsmath}
\usepackage{algorithm}
\usepackage{algorithmic}
\usepackage{multirow}
\def\BibTeX{{\rm B\kern-.05em{\sc i\kern-.025em b}\kern-.08em
    T\kern-.1667em\lower.7ex\hbox{E}\kern-.125emX}}
\begin{document}

\title{Gradient Sparification for Asynchronous Distributed Training\\
\thanks{Identify applicable funding agency here. If none, delete this.}
}

\author{\IEEEauthorblockN{1\textsuperscript{st} Zijie Yan}
\IEEEauthorblockA{\textit{School of Data and Computer Science} \\
\textit{Sun Yat-sen University}\\
Guangzhou, China \\
yanzj@mail2.sysu.edu.cn}

}

\maketitle
\begin{abstract}
Modern large scale machine learning applications require stochastic optimization algorithms to be implemented on distributed computational architectures. A key bottleneck is the communication overhead for exchanging information, such as stochastic gradients, among different nodes. Recently, gradient sparsification techniques have been proposed to reduce communications cost and thus alleviate the network overhead. However, most of gradient sparsification techniques consider only synchronous parallelism and cannot be applied in asynchronous scenarios, such as asynchronous distributed training for federated learning at mobile devices.

In this paper, we present a dual-way gradient sparsification approach (DGS) that is suitable for asynchronous distributed training. We let workers download model difference, instead of the global model, from the server, and the model difference information is also sparsified so that the information exchanged overhead is reduced by sparsifying the dual-way communication between the server and workers. To preserve accuracy under dual-way sparsification, we design a sparsification aware momentum (SAMomentum) to turn sparsification into adaptive batch size between each parameter. We conduct experiments at a cluster of 32 workers, and the results show that, with the same compression ratio but much lower communication cost, our approach can achieve better scalability and generalization ability. 
\end{abstract}

\section{Introduction}
With the increase of training data volume and growing scale of deep neural networks (DNNs), training a large DNNs model may take an impractically long time at a single machine. Distributed training, especially data parallelism, has become essential to reduce the training time of large DNNs model on large data sets \cite{dean2012large,deng2009imagenet,li2014scaling} .
Distributed training relies on distributed optimizers to minimize the objective function of large-scale DNNs. Synchronous stochastic gradient descent (SSGD) \cite{strom2015scalable,coates2013deep} is one of the most popular distributed optimizers, which distributes the workload to multiple workers and aggregates gradients computed by workers into the global model update equivalent to that of single worker but larger batch size training. 
Since SSGD based distributed training may suffer from worker lags, which deteriorates the efficiency and scalability, asynchronous stochastic gradient descent (ASGD) \cite{tsitsiklis1986distributed,recht2011hogwild,zhang2013asynchronous,keuper2015asynchronous} has been proposed to remove synchronization barrier among workers. ASGD is usually realized under the parameter server (PS) architecture \cite{li2014scaling}. PS is a node to collect and aggregate gradients from workers, and workers exchange gradients and model with the server at their own pace. Since there is no longer synchronization among workers, ASGD can significantly speed up the process of distributed training. 

By increasing the number of training nodes and taking advantage of data parallelism, distributed training via SSGD/ASGD can significantly reduce the total computation time of forward-backward passes on the same volume of data. However, either SSGD or ASGD introduces the communication overhead of exchanging model parameters or gradients in each iteration \cite{wangni2018gradient}. 

To cope with the communication challenges in distributed deep learning, quite a number of efforts have been made, and we can either reduce the frequency of communication by increasing the batch size or reduce the data volume of communication in each iteration. Large batch training \cite{goyal2017accurate,wang2017stochastic,jia2018highly} try to scale data-parallelism SGD to more computing nodes without reducing the workload on each node. However, increasing batch size often leads to a significant loss in test accuracy \cite{goyal2017accurate,hoffer2017train}, and sophisticated hyper-parameter tuning like learning rate control \cite{goyal2017accurate,krizhevsky2014one,you2017scaling} must be used to get better convergence accuracy. On the other hand, gradient compression is another powerful method that can largely reduces the volume of exchanged data without affecting convergence performance. 
There are two different ways to realize gradient compression: gradient quantization and gradient sparsification. Gradient quantization, e.g., 1-bit SGD \cite{seide20141}, QSGD \cite{alistarh2016qsgd} and TernGrad \cite{wen2017terngrad}, compress the float-point number with prominent data representation and use fewer bits to represent each value. Gradient sparsification \cite{alistarh2018convergence,stich2018sparsified}, on the other hand, tries to exchange only essential gradient values. The importance of a gradient can be measured by the gradient magnitude or other factors. 
Storm et al. \shortcite{strom2015scalable} prunes gradients using a fixed threshold, while Aji et al. \shortcite{aji2017sparse} and others \cite{chen2018adacomp,dryden2016communication,wangni2018gradient} proposed relative and adaptive thresholds to transmit only the essential gradients. Compare to gradient quantization, gradient sparsification can achieve much higher compression ratio in large scale DNN training. 
However, almost all existing gradient sparsification approaches are designed based on SSGD, i.e., they can be used for only synchronous training. In asynchronous training with ASGD, since workers may be using different model parameters at the same time, they need to download the whole model from server, and compression/sparsification is not applicable. 

In this paper, we propose DGS, a novel approach for asynchronous training to overcome the communication bottleneck by compressing information exchanged. Different from existing asynchronou training, where workers need to download the whole model from the server, we let workers download the model difference between global and local from the server. Accordingly, DGS could sparsify both downward and upward communication to reduce communication volume. Such a dual-way compression approach can significantly reduce communication cost in asynchronous training. More importantly, to avoid loss of accuracy, we design, SAMomentum, a novel momentum suitable for asynchronous training. 
Compared with existing momentum, which can only be used under dense updates, our SAMomentum achieved much better convergence performance in the sparse scenario.

We conducted three empirical studies to evaluate the proposed approach. The experiment results show that our approach has better convergence performance and scalability than existing ones, including ASGD, Gradient Dropping \cite{aji2017sparse}, and Deep Gradient Compression \cite{lin2017deep}. Moreover, our approach works well with a low network bandwidth of 1Gbps, which is significant for asynchronous distributed training in mobile or wireless environments.

The rest of paper is organized as follows: Section 2 discusses related works on distributed training. The preliminaries are described in Section 3. Section 4 presents the design of our dual-way compression approach DGS and the design of the novel momentum SAMomentum. This section also provides proof of the correctness of our design, i.e., with our new momentum, the accuracy of our approach is equivalent to that of enlarging batch size for each model parameter. The experiments and results are reported in Section 5. Finally, we conclude this paper in Section 6.

\section{Related Work}
Researchers have proposed many approaches to optimize the SGD algorithm and communication pattern. The underlying idea is to relax the synchronization restriction to avoid waiting for slow workers. The HOGWILD algorithm \cite{recht2011hogwild} allows workers to read and write global model at will, which has been proven to converge for sparse learning problems. Downpour SGD \cite{dean2012large} extended HOGWILLD to distributed-memory systems, which run multiple minibatches before exchanging gradients so as to reduce communication cost.
 
Another direction is to increase the minibatch size. Traditionally, due to memory constraints and accuracy degradation, minibatch size in deep learning usually less than 256. However, the scaling of data parallelism is limited by the size of minibatch. \cite{goyal2017accurate} proposed warmup approach and linear scaling rule to guarantee the convergence performance. \cite{you2017scaling} further introduce LARS, a method that changes the learning rate independently for each layer based on the norm of their weights and the norm of their gradient. It becomes possible to train with large minibatch sizes like 8k and 32k samples without significant injury on the accuracy, which makes the matrix operations more efficient and reducing the frequency of communication.

Gradient compression approaches, including gradient quantization and sparsification, are proposed to reduce communication data volume. Gradient quantization redeuces the communication overhead by representing gradient values with fewer bits. \cite{gupta2016model} proposed the 16-bit float values representation for model parameters and gradients. 1-Bit SGD \cite{seide20141} and TernGrad \cite{wen2017terngrad} even quantize gradients to binary or ternary values, while still guarantee convergence with marginally reduced accuracy. QSGD \cite{alistarh2016qsgd} randomly quantizes gradients using uniformly distributed quantization points, which also explains the trade-off between model performance and gradient precision. However, even binary gradients can only achieve 32× reduced size, which is not really enough for large models and slow networks. Gradient sparsification approaches try to exchange selected valued rather than all of them. Storm et al. \cite{strom2015scalable} proposed to prune gradients using a static threshold, and got up to 54× speedup with 80 nodes. However, it is hard to determine an appropriate threshold for a neural network in practice. \cite{aji2017sparse} proposed Gradient Dropping, which sends only the top R\% (R is fixed) gradients in terms of size, and accumulates the other gradients locally. \cite{dryden2016communication} proposed to exchange only the important positive and negative gradients, based on their absolute value. DoubleSqueeze-async \cite{tang2019doublesqueeze} performs the compression at both the worker side and the server side. It gathers $m$ gradients at the server like HOGWILD \cite{recht2011hogwild}, and then broadcasts compressed accumulated gradients to all workers. Lin et al. \cite{lin2017deep} proposed momentum correction to correct the disappearance of momentum discounting factor, along with some optimization tricks(including the warmup strategy and gradient clipping), which shows that Top-k sparsification SSGD can converge very closely to SGD.

\section{Preliminary and Motivation}

\subsection{Gradient Sparsification in SSGD}

Various gradient sparsification approaches have been proposed to reduce the communication cost in distributed training. The key idea behind these approaches is to drop part of the stochastic gradient updates and only transmit the rest. For example, Aji et al. \cite{aji2017sparse} propose to sparsify the gradients and transmit the elements with Top-k absolute values. Their sparsification method map the 99\% smallest updates to zero then exchange sparse matrices, which significantly reduce the size of updates with marginally affecting the convergence performance. In order to avoid losing information, gradient sparsification usually accumulate the rest of the gradients locally, eventually, send all of the gradients over time. After each worker contributed the k largest gradients, we need average gradients from all workers than apply the averaged results to each worker. However, the sparsified gradients are generally associated with irregular indices (e.g., COO format), which makes it a challenge to accumulate the selected gradients from all workers efficiently. In decentralized SSGD, recent solutions uses the AllGather collective \cite{renggli2018sparcml}. In parameter-server (PS) based SSGD, the server could do the average operation by adding support of sparse matrix. However, all the above methods to gather gradients are designed for SSGD. In ASGD, since different workers may be installed with different model versions, methods designed for SSGD will no longer work. 
\subsection{Asynchronous SGD}
Same as other SGD algorithms, the goal of asynchronous SGD is to minimize an optimization problem $L\left( \theta \right)$, where $L$ is the objective function, and the vector $\theta$ is the model's parameters. In asynchronous SGD, all $N$ workers compute gradients asynchronously in parallel. After a worker $k$ completes backpropagation with local model
$\theta_{k,\text{prev}\left( k \right)}$, it will send gradients $\nabla L\left( \theta_{k,\text{prev}\left( k \right)} \right)$ to the parameter server and wait for the updated parameter $\theta$ back from server, where $\text{prev}\left( k \right)$ denotes the last iteration that the worker $k$ sent gradient to the server. Once the server receives the gradient $\nabla L\left( \theta_{k,\text{prev}\left( k \right)} \right)$ from the worker $k$, it applies the gradient to its current set of parameters $\theta$, and then sends $\theta$ back to the worker. 

Figure \ref{psarch} shows the overall architecture of existing asynchronous SGD. The upward communications are gradients from worker to the server, and the downward communications are global model parameters from the server to workers.
We can compress upward communication by gradient sparsification methods. However, the downward communications of ASGD are unsuitable for gradient sparsification. This is because, different workers may keep different versions of the model at the same time, so the gradients aggregated at the server is meaningful for only the model version at the server. And if workers download the whole model at downward communications, the network bottleneck still exists. This motivates the proposal of DGS in this paper, a new gradient sparsification approach for ASGD with the sparsification aware momentum.

In DGS, we modify the update operations at the server. Instead of sending the global model to the worker $k$, DGS sends the model difference between global and local, which becomes compressible.
\section{The Proposed Approach DGS}\label{AA}
In this section, we will introduce the detailed design of DGS. Firstly, we describe the dual-way sparsification operations, including the method to track the difference between global model and local model, and operations to do sparsification. Secondly, we present the design of sparsification aware momentum (SAMomentum), which is used to offer a significant optimization boost. At last, we derive the equivalence between DGS and enlarged batch size.
\subsection{Dual-way Gradient Sparsification for Asynchronous Training }\label{two-way-sparsiuxfb01cation-in-asynchronous-sgd}
\textbf{Model Difference Tracking}
In our dual-way gradient sparsification, the server maintains a separate vector $v^{k}$ for each worker, which is the accumulation of the gradients that have been sent to
worker $k$. The server no longer maintains the global model but maintains the
accumulation of updates $M_{t}$. In the following, for simplicity of presentation, we denote the current stochastic gradient $\nabla L\left( \theta_{k,t\left( k \right)} \right)$ by $\nabla_{k,t}$ for short:
\begin{equation}
M_{t + 1} = M_{t} - \eta\nabla_{k,\text{prev}}
\end{equation}
$\eta$ is the learning rate, and $t$ is the scalar timestamp
that tracks the number of updates made to the server parameters ($t$ starts at 0 and is incremented by one for each update).
$M_{t}$ is the difference between the initial model and the global model, and $M_{0}$ is a zero vector:
\begin{equation}
M_{t + 1} = \theta_{t + 1} - \theta_{0}
\end{equation}
After updating $M$, the server will send $G_{k,t + 1}$ to
the worker $k$ and add $G_{k,t + 1}$ on
$v_{k,\text{prev}\left( k \right)}\,$:
\begin{equation}
G_{k,t + 1} = M_{t + 1} - v_{k,\text{prev}\left( k \right)}
\end{equation}
\begin{equation}
\begin{split}
v_{k,t + 1}\, &= v_{k,\text{prev}\left( k \right)}\, + \, G_{k,t + 1}\\ 
&= v_{k,\text{prev}\left( k \right)} + M_{t + 1} - v_{k,\text{prev}\left( k \right)} \\
&=M_{t + 1}    
\end{split}
\end{equation}

In ASGD, a worker $k$ receives the global model $\theta_{t + 1}$ from server, replaces its local model $\theta_{k,\text{prev}\left( k \right)}$ with
$\theta_{t + 1}$, and then moves to next iteration. However, DGS chooses to transmits
$G_{k,t + 1}$ rather than the global model:

\begin{equation}
\begin{split}
\theta_{k,t + 1} &= \theta_{k,\text{prev}\left( k \right)} + G_{k,t + 1} \\
&= {\theta_{0} + M}_{\text{prev}\left( k \right)} + M_{t + 1} - v_{k,\text{prev}\left( k \right)} \\
&= \theta_{0} + M_{\text{prev}\left( k \right)} + M_{t + 1} - M_{\text{prev}\left( k \right)} \\
&= \theta_{0} + M_{t + 1} = \theta_{t + 1}
\end{split}\label{eq1}
\end{equation}

Eq.\eqref{eq1} indicates that DGS without sparsification is equivalent to ASGD. DGS transmits the difference between the global model and local model $G_{k,t + 1}$, which can be sparsely compressed. 

\textbf{Sparsification Operations} The following pseudo-code describes how to perform dual-way gradient sparsification in DGS.
Algorithm \ref{alg1} shows the gradient dropping scheme used at the workers in DGS, which is quite similar to the Top-k sparsification in distributed SGD \cite{aji2017sparse}. Algorithm \ref{alg2} shows the update rules at the server. Please notice that, at line 5 of Algorithm \ref{alg2}, there is a switch for secondary compression. Normally, there is no secondary compression for $G_{k,t + 1}$ at the server, because $G_{k,t + 1}$ is the accumulation of several sparse updates and $G_{k,t + 1}$ itself is highly sparse. However, for the environments with very limited communication resources (e.g., mobile devices) or a very large number of workers, secondary compression (line 5-11 of Algorithm \ref{alg2}) can be included to further reduce data exchanged.

\textbf{Secondary compression} DGS enables the ability to performs secondary compression on the server. Substituting
$M_{t + 1} - v_{k,\text{prev}\left( k \right)}$with
$s\text{parse}\left( M_{t + 1} - v_{k,\text{prev}\left( k \right)} \right)$
yields the update rule of secondary compression:
\begin{subequations} \label{eq6}
	\begin{gather}
        G_{k,t + 1} = sparse\left( M_{t + 1} - v_{k,\text{prev}\left( k \right)} \right)\\
        v_{k,t + 1}\, = v_{k,\text{prev}\left( k \right)}\, + \, G_{k,t + 1}\,
	\end{gather}
\end{subequations}
The server implicitly accumulates remaining gradient locally. Eventually, these gradients become large enough to be transmitted immediately.
Lines 5-11 of algorithm \ref{alg2} show how the server compresses $G_{k,t + 1}$ in secondary compression, which eliminates the overhead of the downward communication. 


\begin{algorithm}
  \caption{DGS on worker $k$}
  \label{alg1}
  \begin{algorithmic}[1]
  \REQUIRE  Dataset $\mathcal{X}$
  \REQUIRE  Initial parameters $\theta_{0}=\{theta[0],...,theta[J]\}$
  \REQUIRE  optimization function $SGD$
  \REQUIRE  $encode()$ function pack nonzero gradients to  coordinate format. 
  \REQUIRE $decode()$ function unpack nonzero gradients from coordinate format.
  \STATE $\theta_{k,0}$ = $\theta_{0}$
  \STATE $v_k \gets \{0,...,0\}$
  \FOR{$t=0,1,...$}
  \STATE Sample data $x$ from $\mathcal{X}$
  \STATE $\nabla_{k,t+1}$ $\gets Backward(x,\theta_t) $
  \STATE $v_{k,t+1} \gets v_{k,t}+\eta\nabla_{k,t+1}$
  \FOR{ $j=0,...,J$}
  \STATE $thr \gets R\% \text{ of } \left|v_{k,t+1}^{}[j]\right|$
  \STATE $Mask \gets \left|v_{k,t+1}^{}[j]\right|>thr$
  \STATE $v_{k,t+1}^{}[j] \gets v_{k,t+1}^{}[j] \odot \neg Mask$
  \STATE $g_{k,t+1}^{}[j] \gets v_{k,t+1}^{}[j] \odot Mask$
  \ENDFOR
  \STATE Send $encode(g_{k,t+1}^{})$ to the server
  \STATE Recieve $G_{k,t+1}$ from the server
  \STATE $\theta_{t+1} \leftarrow SGD\left(\theta_{t}, decode(G_{k,t+1})\right)$
  \ENDFOR
  \end{algorithmic}
\end{algorithm}

\begin{algorithm}
  \caption{DGS on server}
  \label{alg2}
  \begin{algorithmic}[1]
  \REQUIRE  Initial parameters $\theta_{0}=\{theta[0],...,theta[J]\}$
  \REQUIRE  $encode()$ function and $decode()$ function
  \STATE $M_{t+1}^k \gets \{0,...,0\}$
  \WHILE{ Receive encoded $g_{}^{k}$ from worker $k$ }
  \STATE $M_{t+1}^k \gets M_t^k- decode(g_{}^{k})$
  \STATE $G_{k,t + 1} \gets M_{t + 1} - v_{k,\text{prev}\left( k \right)}$
  \IF{Need secondary compression} 
        \FOR{ $j=0,...,J$}
        \STATE $thr \gets R\% \text{ of } \left|G_{k,t + 1}[j]\right|$
        \STATE $Mask \gets \left|G_{k,t + 1}[j]\right|>thr$
        \STATE $G_{k,t + 1}[j] \gets G_{k,t + 1}[j] \odot  Mask$
        \ENDFOR
　\ENDIF
　\STATE Send $encode(G_{k,t + 1})$ to the worker $k$
　\STATE $v_{k,t+1} \gets v_{k,\text{prev}\left( k \right)}-G_{k,t + 1}$
　\STATE $\text{prev}(k)=t+1$
  \ENDWHILE
  
  \end{algorithmic}
\end{algorithm}
\subsection{Sparsification Aware Momentum}\label{momentum}

Momentum \cite{polyak1992acceleration} is ubiquitous in deep learning training, as it is known to offer a significant optimization boost. Momentum for SSGD training can be calculated as follows:

\begin{equation}
u_{t} = mu_{t - 1} + \eta\nabla_{t},\,\,\theta_{t + 1} = \theta_{t} - u_{t}
\end{equation}
$u_{t}$ is the velocity. On parameter server based ASGD with $N$ nodes, it becomes:
\begin{equation}
u_{t} = mu_{t - 1} + \eta\nabla_{k,\text{prev}},\,\,\theta_{t + 1} = \theta_{t} - u_{t}
\end{equation}

With gradients sparsiﬁcation as in Algorithm \ref{alg1}, it further changes to be \cite{lin2017deep}:

\begin{equation}
\label{eq9}
g_{k,t} = \eta*sparsify\left(\nabla_{k,\text{prev}} \right)
\end{equation}
\begin{equation}
u_{t} = mu_{t - 1} + g_{k,t},\text{\ \ }\theta_{t + 1} = \theta_{t} - u_{t}
\end{equation}

The function $sparsify\left(\right)$ will zero out gradients less than the threshold $thr$ and the function $unsparsify\left(\right)$ will zero out gradients lager than the threshold. We name the result of $unsparsify\left(\nabla_{k,\text{prev}}\right)$ as remaining gradients. 
Remaining gradients \textbf{will not} participate in momentum update in Eq. \eqref{eq9} since workers have not sent them yet, which results in broken momentum and consequently, loss of convergence \cite{lin2017deep}.

Sparsification Aware Momentum (SAMomentum) is a novel momentum designed for gradient sparsification scenario. DGS accumulates SAMomentum locally at each worker instead of collecting it at the server, and rescales remaining gradients in $u_{t}$:
\begin{subequations} \label{updaterule}
	\begin{gather}
        u_{k,t} = mu_{k,\text{prev}\left( k \right)} + \eta\nabla_{k,\text{prev}}  + \notag \\
        unsparsify\left( mu_{k,\text{prev}\left( k \right)} + \eta\nabla_{k,\text{prev}} \right)*\left(\frac{1}{m}-1 \right)\\
        g_{k,t} = sparsify\left( mu_{k,\text{prev}\left( k \right)} + \eta\nabla_{k,\text{prev}} \right)\\
        \theta_{t + 1} = \theta_{t} - g_{k,t}
	\end{gather}
\end{subequations}
$u_{k,c}^{\left(i\right)}$ denote the i-th position of a flattened velocity $u$ at the worker $k$ with local timestamp $c$, where $c$ is the $c$-th iteration of the worker. $T$ is the length of the sparse update interval between two iterations during which $u_{k,t}^i$ is sent. After each step, $u_{k,c}^{\left(i\right)}$ equals to:
\begin{equation}
u_{k, c}^{(i)}=\left\{\begin{array}{ll}{m u_{k, c-1}^{(i)}+\eta \nabla_{k, c}^{(i)}} & {>thr} \\ {\left(m u_{k, c-1}^{(i)}+\eta \nabla_{k, c}^{(i)}\right) *\frac{1}{m}} & {\leq thr}\end{array}\right.
\end{equation}
Workers send encoded $u_{k,t+1}^{}[j] \odot Mask$ to the server, as shown in Algorithm \ref{alg3}.
\begin{algorithm}
  \caption{DGS on worker $k$ with Sparsification Aware momentum}
  \label{alg3}
  \begin{algorithmic}[1]
  \REQUIRE  Dataset $\mathcal{X}$
  \REQUIRE  Initial parameters $\theta_{0}=\{theta[0],...,theta[J]\}$
  \REQUIRE  optimization function $SGD$
  \REQUIRE  $encode()$ function and $decode()$ function
  \STATE $\theta_{k,0}$ = $\theta_{0}$
  \STATE $u_k \gets \{0,...,0\}$
  \FOR{$t=0,1,...$}
  \STATE Sample data $x$ from $\mathcal{X}$
  \STATE $\nabla_{k,t+1}$ $\gets Backward(x,\theta_t) $
  \STATE $u_{k,t+1} \gets mu_{k,t}+\eta\nabla_{k,t+1}$
  \FOR{ $j=0,...,J$}
  \STATE $thr \gets R\% \text{ of } \left|u_{k,t+1}^{}[j]\right|$
  \STATE $Mask \gets \left|u_{k,t+1}^{}[j]\right|>thr$
  \STATE $g_{k,t+1}^{}[j] \gets u_{k,t+1}^{}[j] \odot Mask$
  \STATE $u_{k,t+1}^{}[j] \gets u_{k,t+1}^{}[j] + \left(\frac{1}{m}-1 \right) u_{k,t+1}^{}[j] \odot \neg Mask$
  \ENDFOR
  \STATE Send $encode(g_{k,t+1}^{})$ to the server
  \STATE Receive $G_{k,t+1}$ from the server
  \STATE $\theta_{t+1} \leftarrow SGD\left(\theta_{t}, decode(G_{k,t+1})\right)$
  \ENDFOR
  \end{algorithmic}
\end{algorithm}

\subsection{Equivalence between DGS and Enlarged Batch Size}\label{momentum}
Suppose $u_{k}^{\left(i\right)}$ is sent to the server at $c$ and $c+T$, therefore $u_{k}^{\left(i\right)}$ is smaller than $thr$ between time $c+1$ and $c+T-1$, then greater than $thr$ at time $c+T$. The change of velocity value  $u_{k}^{\left(i\right)}$ equals to $\eta \sum_{i=1}^{T} \nabla_{k, c+i}^{(i)}$:
\begin{equation}
\begin{split}
u_{k, c+T}^{(i)}&=m u_{k, c+T-1}^{(i)}+\eta \nabla_{k, c+T}^{(i)} \\
&=m\left(\left(m u_{k, c+T-2}^{(i)}+\eta \nabla_{k, c+T-1}^{(i)}\right) * \frac{1}{m}\right)+\eta \nabla_{k, c+T}^{(i)} \\
&= m u_{k, c+T-2}^{(i)} + \eta \nabla_{k, c+T-1}^{(i)} +\eta \nabla_{k, c+T}^{(i)}  \\
&=\cdots\\
&=m u_{k, c}^{(i)}+\eta \sum_{i=1}^{T} \nabla_{k, c+i}^{(i)}
\end{split}\label{wodetuidao}
\end{equation}
which can be considered as vanilla momentum SGD (MSGD) increasing batch size and learning rate by $T$ times. With increasing batch size and learning rate, vanilla MSGD becomes:
\begin{equation}
\begin{split}
u_{k, c+T}^{(i)}&=m u_{k, c}^{(i)} + T\eta*\frac{1}{T}\left( \nabla_{k, c+1}^{(i)} +\cdots +\nabla_{k, c+T}^{(i)}\right) \\
&=m u_{k, c}^{(i)}+\eta \sum_{i=1}^{T} \nabla_{k, c+i}^{(i)}
\end{split}\label{increasebatchsize}
\end{equation}

For every single parameter of weight $\theta$, Eq. \eqref{wodetuidao} is equivalent to \eqref{increasebatchsize}. The underlying idea of \eqref{updaterule} is that, SAMomentum adaptively enlarge the batch size for every single parameter without introducing any hyperparameters. Note that, DGS with SAMomentum does not need local gradient accumulation, as shown in Algorithm \ref{alg3}, which is necessary for DGC and other gradient sparsification approaches. Now, the momentum in Eq. \eqref{updaterule} is the one used in our design, which can save lots of memory compared with DGC. In other words, we basically turn the sparsification into the magnification of batch size. 

Recent research like \cite{goyal2017accurate},\cite{you2017scaling} attempted to enlarge the batch size of the entire model for efficient training, which makes it possible to train DNNs with large batch size without significant loss of accuracy. We also enlarge the batch size in distributed training, but our approach is in the parameter level rather than model level.
What's more, different from the update in existing works as shown in Eq. \eqref{increasebatchsize}, during the sparse update interval, DGS continuously receives updates for the parameter $\theta_{k}^{\left(i\right)}$, which can obviously decrease the asynchronous staleness of $u_{k}^{\left(i\right)}$.

In SSGD and ASGD, each parameter of the local model has the same and fixed update interval and batch size. However, sparsification techniques like Gradient Dropping introduced different update intervals to each parameter, since workers only send part of gradients in each iteration. This change makes each parameter have their own asynchronous update pace. SAMomentum takes advantage of such a change, and applies adaptive batch size in element-wise based on, so as to avoid information losing in sparse asynchronous training with momentum, in spite of DGS with SAMomentum do not accumulate residual gradients (the $v_{k,t}$ in Algorithm \ref{alg1}) anymore.

\section{Experimental Evaluation}
The evaluation is conducted using a 36-GPU cluster, with different neural network models and datasets. We examine the performance via two types of deep learning tasks: image classification and speech recognition. We also compare with four other approaches：MSGD, ASGD, Gradient Dropping, and DGC. ASGD is vanilla asynchronous SGD without gradient sparsification and MSGD is a single-node momentum SGD. 

However, Gradient Dropping and DGC is originally designed based on SSGD, and do not work in asynchronous training. Therefore, for comparison purpose, We implemented an asynchronous version of Gradient Dropping and DGC by adding model difference based compression as in our DGS, and they are denoted as GD-async and DGC-async in experiments.  
\subsection{Dataset and Models}
Image Classification: We use ResNet-18 and ResNet-50 \cite{he2016deep} on Cifar10 dataset. Cifar10 consists of 50,000 training images and 10,000 validation images in 10 classes \cite{krizhevsky2009learning}. The baseline training using vanilla MSGD with a momentum of 0.7. The momentum coefficient of our approach and DGC-async is 0.7, too. All experiments decrease the learning rate by the factor of 0.1 at epoch 30 and 40 out of 50 epochs. To simplify comparisons, we do not include other training tricks for improving accuracy.

Speech Recognition: The AN4 dataset contains 948 training and 130 test utterances. We train a 5-layer LSTM with 800 hidden units, and the hyperparameters settings are: epochs, $100$; learning rate, $4*10^-4$; weight decay, $1.25*10^{-5}$; momentum, $0.7$; learning rate anneal, $1.01$.
\subsection{Experiments Setup}
Hardware: the distributed environment is configured as a 32-GPU cluster with eight machines, each of which is equipped with 2 Intel Xeon E5-2660-V3 CPUs and 4 NVIDIA Tesla K80 GPUs. The default network between workers and the server is 10 Gbps Ethernet. In all of our experiments, we use each GPU as a worker.

Software: All GPU machines are installed with Linux 3.10.0, NVIDIA GPU driver 390.30 and CUDA 8. We implement all the algorithms via PyTorch0.4, a popular lightweight distributed deep learning framework with great flexibility. The parameter server is implemented using PyTorch distributed API with TCP backend. Furthermore, only vanilla MSGD is training with a single node, and others all execute asynchronously based on our model difference compression as in (Algorithm \ref{alg1},\ref{alg2}), so the result might be different from their synchronous experimental results.

\subsection{Results and Analysis}
\begin{figure}[htbp]
    \centering
    \includegraphics[width=.95\linewidth,height=.35\textheight,keepaspectratio]{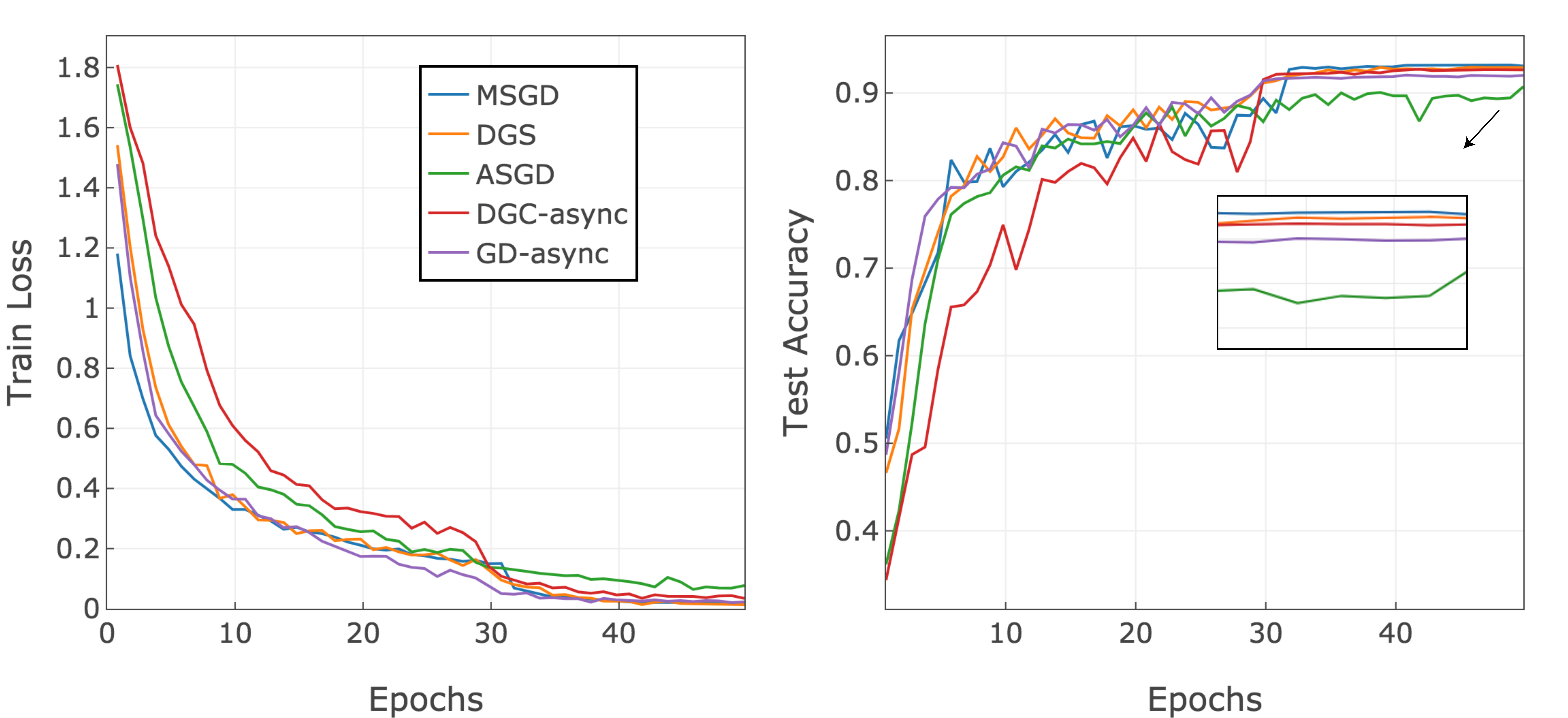}
    \caption{Training ResNet-18 on Cifar10}
    \label{fig1}
\end{figure}

\paragraph{Image Classiﬁcation} We first examine our approach on Cifar10 dataset. Figure \ref{fig1} is the Top-1 accuracy and training loss of ResNet-18 on Cifar10 with 4 workers. The gradient sparsity of DGC-async, GD-async, and DGS is 99\%. The learning curve of GD-async (purple) and ASGD (green) is worse than MSGD (blue) due to gradient staleness. With momentum correction, the learning curve of DGC-async (red) converges slightly slower, but the accuracy is much closer to the baseline. DGS outperforms the other three approaches, and its convergence performance is even close to single-node MSGD. Moreover, the accuracy curve of DGS converges smoothly and stably, which is obviously better than other approaches, especially DGC. Table \ref{table1} shows the detailed accuracy results. The accuracy of ResNet-18 converges very well using our distributed approach with 4  workers.

\begin{table}[htbp]
\caption{Results of ResNet-18 trained on Cifar10}
\centering
\resizebox{.95\columnwidth}{!}{
\smallskip\begin{tabular}{|c|c|c|c|} 
\hline
Model                      & Training Method   & Workers in total & Accuracy  \\ 
\hline
\multirow{5}{*}{ResNet-18} & MSGD               & 1                & 93.08\%   \\ 
\cline{2-4}
                           & ASGD              & 4                & 90.74\%   \\ 
\cline{2-4}
                           & GD-async & 4                & 92.01\%   \\ 
\cline{2-4}
                           & DGC-async               & 4                & 92.64\%   \\ 
\cline{2-4}
                           & Our approach     & 4                & 92.91\%   \\
\hline

\end{tabular}
}
\label{table1}
\end{table}

\paragraph{Speech Recognition}
For speech recognition, Table \ref{table2} shows the average word error rate (WER) of a 5-layer LSTM on AN4 Dataset. GD-async and ASGD not converge on 4 worksers. The results show that our DGS achieves the same improvement as that for the image network.
\begin{table}[htbp]
\caption{Results of 5-Layer LSTM on AN4}
\centering
\resizebox{.98\columnwidth}{!}{
\smallskip
\begin{tabular}{|c|c|c|c|c|}
\hline
Model                         & Training Method & Workers & Batch Size & Average WER \\ \hline
\multirow{3}{*}{5-layer LSTM} & SGD             & 1                & 20         & 26.2\%                         \\ \cline{2-5} 
                              & DGC-Async       & 4                & 5          & 23.54\%                        \\ \cline{2-5} 
                              & DGS             & 4                & 5          & 21.51\%                        \\ \hline
\end{tabular}
}
\label{table2}
\end{table}
\subsection{Scalability and Generalization Ability}

In this experiment, we run our approach, ASGD, GD-async, and DGC-async on 1, 4, 8, 16, 32 workers asynchronously, and compare their test accuracy. All experiments have the same hyperparameter setting, except the batch size. The baseline approach single-node MSGD runs using a mini-batch size of 256, resulting in a test accuracy of 93.08\%. 

In Table \ref{table2}, we can observe that, the test accuracy of other approach decreases as the number of workers increases. This is because, with more nodes, the more staleness asynchrony brings. However, the test accuracy of our approach in 4, 8 and 16 workers is 92.91\%, 93.32\%, and 92.98\% respectively. Therefore, our approach has better converge performance and even defeats the staleness brought by asynchronous in distributed scenarios. Compare to other approaches on 32 workers, our approach achieves the best accuracy, and the accuracy only drops a little (-0.39\%) due to a large number of workers. At the same time, the convergence performance of other methods is greatly reduced: ASGD drops to 88.36\% (-4.71\%), GD-async drops to 91\% (-2.08\%) and DGC-async drops to 91.86\% (-1.22\%). Experiments results above show that, our approach scales very well when the number of workers increases and does not negatively affect (or perhaps helps) generalization.

\begin{figure}[htbp]
    \centering
    \includegraphics[width=.95\linewidth,height=.25\textheight,keepaspectratio]{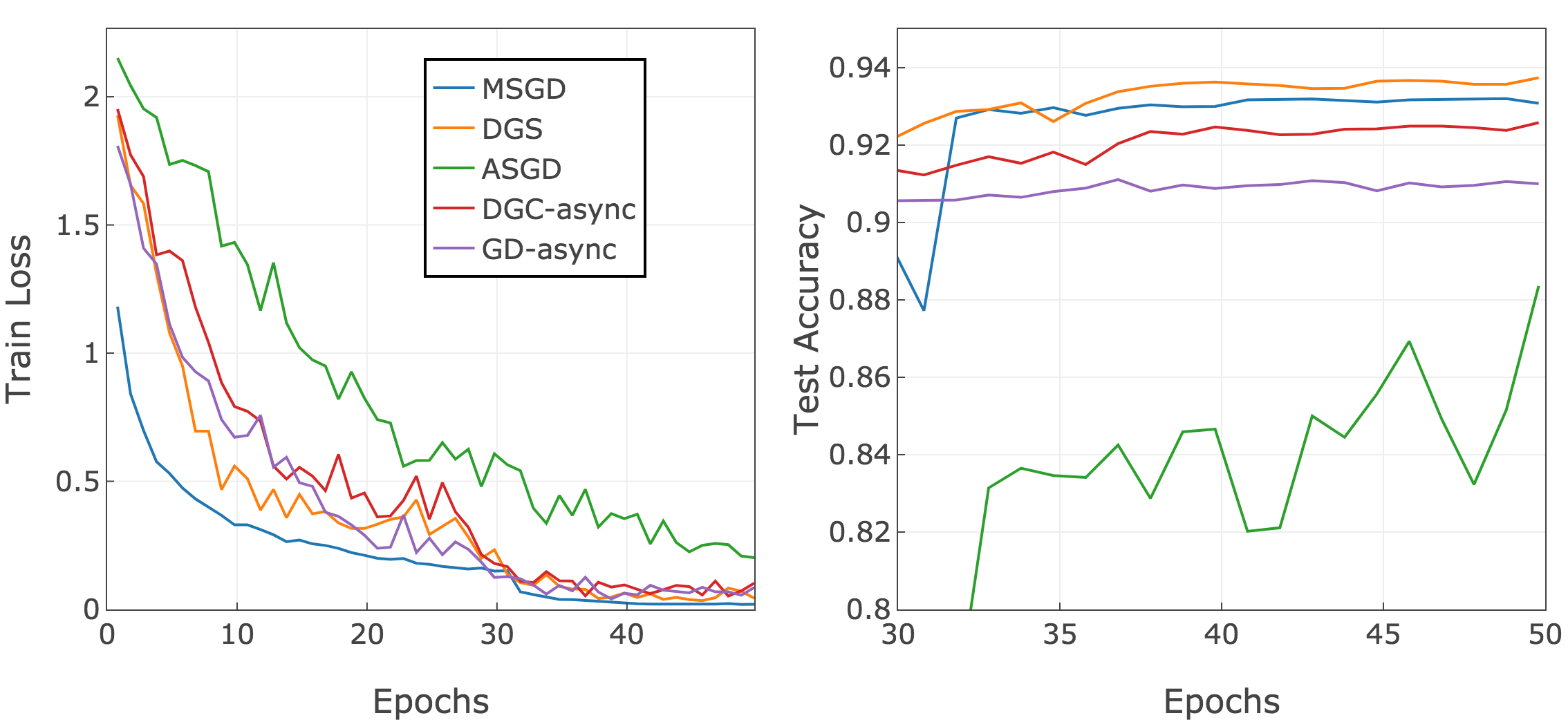}
    \caption{Tuned ResNet-18 on 32 workers}
    \label{fig3}
\end{figure}

\begin{figure}[htbp]
    \centering
    \includegraphics[width=.95\linewidth,height=.2\textheight,keepaspectratio]{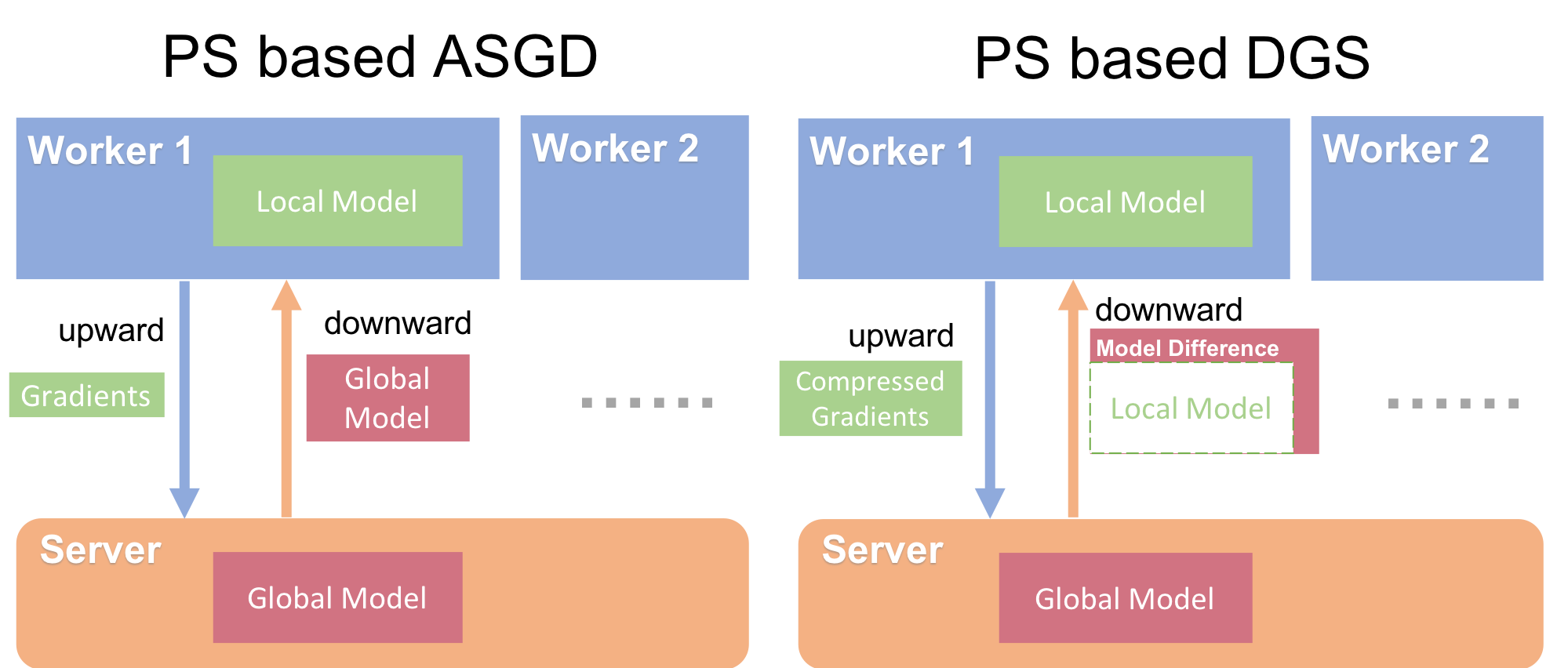}
    \caption{Architecture of PS Based Asynchronous Training}
    \label{psarch}
\end{figure}

Further, we have got amazing results by changing hyperparameters. Since asynchrony introduces momentum to the SGD update \cite{mitliagkas2016asynchrony}, we reduced the momentum from 0.7 to 0.3 on 32 workers. Surprisingly, the test accuracy increased to 93.7\%. Figure \ref{fig3} shows that our approach (yellow) closely follows the curve of single node MSGD (blue) and achieves better accuracy eventually. 

\begin{table}[htbp]
\caption{ResNet-18 trained on Cifar10 Dataset}
\centering
\resizebox{.95\columnwidth}{!}{
\smallskip
\begin{tabular}{|c|c|c|cc|} 
\hline
Workers in total    & \multicolumn{1}{l|}{Batchsize per iteration}                    & Training Method   & \multicolumn{2}{c|}{Accuracy}        \\ 
\hline
\multirow{5}{*}{1}  & \multirow{5}{*}{\begin{tabular}[c]{@{}c@{}}256\\ \end{tabular}} & MSGD (beaseline)  & 93.08\%          & -                 \\ 
\cline{3-5}
                    &                                                                 & ASGD              & 91.54\%          & -1.54\%           \\ 
\cline{3-5}
                    &                                                                 & Gradient Dropping & 92.15\%          & -0.93\%           \\ 
\cline{3-5}
                    &                                                                 & DGC               & 92.75\%          & -0.33\%           \\ 
\cline{3-5}
                    &                                                                 & Our apporoach     & \textbf{92.97\%} & \textbf{-0.11\%}  \\ 
\hline
\multirow{4}{*}{4}  & \multirow{4}{*}{128}                                            & ASGD              & 90.7\%           & -2.38\%           \\ 
\cline{3-5}
                    &                                                                 & Gradient Dropping & 92.01\%          & -1.07\%           \\ 
\cline{3-5}
                    &                                                                 & DGC               & 92.64\%          & -0.44\%           \\ 
\cline{3-5}
                    &                                                                 & Our apporoach     & \textbf{92.91\%} & \textbf{-0.17\%}  \\ 
\hline
\multirow{4}{*}{8}  & \multirow{4}{*}{64}                                             & ASGD              & 90.46\%          & -2.62\%           \\ 
\cline{3-5}
                    &                                                                 & Gradient Dropping & 91.81\%          & -1.27\%           \\ 
\cline{3-5}
                    &                                                                 & DGC               & 92.37\%          & -0.71\%           \\ 
\cline{3-5}
                    &                                                                 & Our apporoach     & \textbf{93.32\%} & \textbf{+0.24\%}  \\ 
\hline
\multirow{4}{*}{16} & \multirow{4}{*}{32}                                             & ASGD              & 90.53\%          & -3.01\%           \\ 
\cline{3-5}
                    &                                                                 & Gradient Dropping & 91.43\%          & -1.65\%           \\ 
\cline{3-5}
                    &                                                                 & DGC               & 92.28\%          & -0.80\%           \\ 
\cline{3-5}
                    &                                                                 & Our apporoach     & \textbf{92.98\%} & \textbf{-0.10\%}  \\ 
\hline
\multirow{4}{*}{32} & \multirow{4}{*}{16}                                             & ASGD              & 88.36\%          & -4.71\%           \\ 
\cline{3-5}
                    &                                                                 & Gradient Dropping & 91\%             & -2.08\%           \\ 
\cline{3-5}
                    &                                                                 & DGC               & 91.86\%          & -1.22\%           \\ 
\cline{3-5}
                    &                                                                 & Our apporoach     & \textbf{92.69\%} & \textbf{-0.39\%}  \\
\hline
\end{tabular}
}
\label{table3}
\end{table}
\subsection{The Effect of Dual-way Sparsification with Low bandwidth}
\begin{figure}[htbp]
    \centering
    \includegraphics[width=.95\linewidth,height=.2\textheight,keepaspectratio]{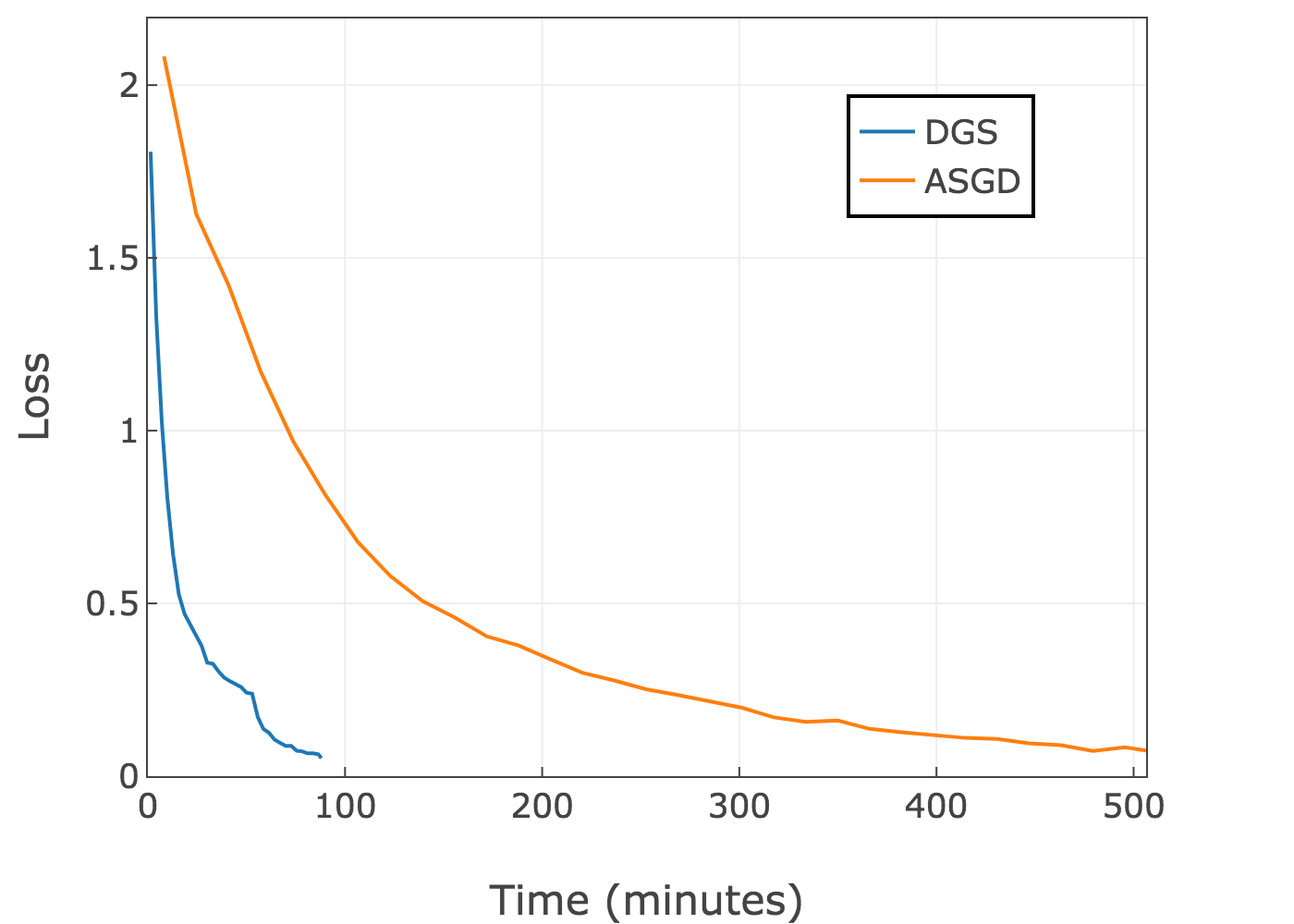}
    \caption{Time vs Training Loss on 8 workers with 1Gbps Ethernet}
    \label{fig4}
\end{figure}

We train ResNet18 of 50 epochs on 8 workers using ASGD and DGS respectively. The network is 1GB Ethernet, and the compress ratio of the secondary compression is 99\%. Comparing ASGD and DGS in Figure \ref{fig4}, we can find that DGS benefits a lot from Dual-way Sparsification. Our approach completes the training for 88 minutes, while ASGD takes 506 minutes, resulting in a speedup of $5.7\times$. 
\section{Conclusion and Future Work}
We propose a novel sparsification approach DGS for asynchronous distributed training. Its major novelty lies in the dual-way compression, which is enabled the delicately designed model difference tracking method. To avoid slowdown in convergence, we also design the sparsification aware momentum (SAMomentum) to transform sparsification into adaptively enlarge batch size, so as to bring significant optimization boost. DGS enables large-scale asynchronous distributed training with inexpensive, commodity networking infrastructure.

In future, the combination of DGS and other compression approaches (e.g. TernGrad \shortcite{wen2017terngrad}, randomly coordinates dropping \shortcite{wangni2018gradient}) can be considered. 
Also, the new momentum SAMomentum is a general design and can be used to design new synchronization training approaches.

\bibliography{bibfile1}

\begin{thebibliography}{}

\bibitem[\protect\citeauthoryear{Aji and Heafield}{2017}]{aji2017sparse}
Aji, A.~F., and Heafield, K.
\newblock 2017.
\newblock Sparse communication for distributed gradient descent.
\newblock {\em arXiv preprint arXiv:1704.05021}.

\bibitem[\protect\citeauthoryear{Alistarh \bgroup et al\mbox.\egroup
  }{2016}]{alistarh2016qsgd}
Alistarh, D.; Li, J.; Tomioka, R.; and Vojnovic, M.
\newblock 2016.
\newblock Qsgd: Randomized quantization for communication-optimal stochastic
  gradient descent.
\newblock {\em arXiv preprint arXiv:1610.02132}.

\bibitem[\protect\citeauthoryear{Alistarh \bgroup et al\mbox.\egroup
  }{2018}]{alistarh2018convergence}
Alistarh, D.; Hoefler, T.; Johansson, M.; Konstantinov, N.; Khirirat, S.; and
  Renggli, C.
\newblock 2018.
\newblock The convergence of sparsified gradient methods.
\newblock In {\em Advances in Neural Information Processing Systems},
  5973--5983.

\bibitem[\protect\citeauthoryear{Chen \bgroup et al\mbox.\egroup
  }{2018}]{chen2018adacomp}
Chen, C.-Y.; Choi, J.; Brand, D.; Agrawal, A.; Zhang, W.; and Gopalakrishnan,
  K.
\newblock 2018.
\newblock Adacomp: Adaptive residual gradient compression for data-parallel
  distributed training.
\newblock In {\em Thirty-Second AAAI Conference on Artificial Intelligence}.

\bibitem[\protect\citeauthoryear{Coates \bgroup et al\mbox.\egroup
  }{2013}]{coates2013deep}
Coates, A.; Huval, B.; Wang, T.; Wu, D.; Catanzaro, B.; and Andrew, N.
\newblock 2013.
\newblock Deep learning with cots hpc systems.
\newblock In {\em International conference on machine learning},  1337--1345.

\bibitem[\protect\citeauthoryear{Dean \bgroup et al\mbox.\egroup
  }{2012}]{dean2012large}
Dean, J.; Corrado, G.; Monga, R.; Chen, K.; Devin, M.; Mao, M.; Senior, A.;
  Tucker, P.; Yang, K.; Le, Q.~V.; et~al.
\newblock 2012.
\newblock Large scale distributed deep networks.
\newblock In {\em Advances in neural information processing systems},
  1223--1231.

\bibitem[\protect\citeauthoryear{Deng \bgroup et al\mbox.\egroup
  }{2009}]{deng2009imagenet}
Deng, J.; Dong, W.; Socher, R.; Li, L.-J.; Li, K.; and Fei-Fei, L.
\newblock 2009.
\newblock Imagenet: A large-scale hierarchical image database.
\newblock In {\em 2009 IEEE conference on computer vision and pattern
  recognition},  248--255.
\newblock Ieee.

\bibitem[\protect\citeauthoryear{Dryden \bgroup et al\mbox.\egroup
  }{2016}]{dryden2016communication}
Dryden, N.; Moon, T.; Jacobs, S.~A.; and Van~Essen, B.
\newblock 2016.
\newblock Communication quantization for data-parallel training of deep neural
  networks.
\newblock In {\em 2016 2nd Workshop on Machine Learning in HPC Environments
  (MLHPC)},  1--8.
\newblock IEEE.

\bibitem[\protect\citeauthoryear{Goyal \bgroup et al\mbox.\egroup
  }{2017}]{goyal2017accurate}
Goyal, P.; Doll{\'a}r, P.; Girshick, R.; Noordhuis, P.; Wesolowski, L.; Kyrola,
  A.; Tulloch, A.; Jia, Y.; and He, K.
\newblock 2017.
\newblock Accurate, large minibatch sgd: Training imagenet in 1 hour.
\newblock {\em arXiv preprint arXiv:1706.02677}.

\bibitem[\protect\citeauthoryear{Gupta, Zhang, and Wang}{2016}]{gupta2016model}
Gupta, S.; Zhang, W.; and Wang, F.
\newblock 2016.
\newblock Model accuracy and runtime tradeoff in distributed deep learning: A
  systematic study.
\newblock In {\em 2016 IEEE 16th International Conference on Data Mining
  (ICDM)},  171--180.
\newblock IEEE.

\bibitem[\protect\citeauthoryear{He \bgroup et al\mbox.\egroup
  }{2016}]{he2016deep}
He, K.; Zhang, X.; Ren, S.; and Sun, J.
\newblock 2016.
\newblock Deep residual learning for image recognition.
\newblock In {\em Proceedings of the IEEE conference on computer vision and
  pattern recognition},  770--778.

\bibitem[\protect\citeauthoryear{Hoffer, Hubara, and
  Soudry}{2017}]{hoffer2017train}
Hoffer, E.; Hubara, I.; and Soudry, D.
\newblock 2017.
\newblock Train longer, generalize better: closing the generalization gap in
  large batch training of neural networks.
\newblock In {\em Advances in Neural Information Processing Systems},
  1731--1741.

\bibitem[\protect\citeauthoryear{Jia \bgroup et al\mbox.\egroup
  }{2018}]{jia2018highly}
Jia, X.; Song, S.; He, W.; Wang, Y.; Rong, H.; Zhou, F.; Xie, L.; Guo, Z.;
  Yang, Y.; Yu, L.; et~al.
\newblock 2018.
\newblock Highly scalable deep learning training system with mixed-precision:
  Training imagenet in four minutes.
\newblock {\em arXiv preprint arXiv:1807.11205}.

\bibitem[\protect\citeauthoryear{Keuper and
  Pfreundt}{2015}]{keuper2015asynchronous}
Keuper, J., and Pfreundt, F.-J.
\newblock 2015.
\newblock Asynchronous parallel stochastic gradient descent: A numeric core for
  scalable distributed machine learning algorithms.
\newblock In {\em Proceedings of the Workshop on Machine Learning in
  High-Performance Computing Environments}, ~1.
\newblock ACM.

\bibitem[\protect\citeauthoryear{Krizhevsky, Hinton, and
  others}{2009}]{krizhevsky2009learning}
Krizhevsky, A.; Hinton, G.; et~al.
\newblock 2009.
\newblock Learning multiple layers of features from tiny images.
\newblock Technical report, Citeseer.

\bibitem[\protect\citeauthoryear{Krizhevsky}{2014}]{krizhevsky2014one}
Krizhevsky, A.
\newblock 2014.
\newblock One weird trick for parallelizing convolutional neural networks.
\newblock {\em arXiv preprint arXiv:1404.5997}.

\bibitem[\protect\citeauthoryear{Li \bgroup et al\mbox.\egroup
  }{2014}]{li2014scaling}
Li, M.; Andersen, D.~G.; Park, J.~W.; Smola, A.~J.; Ahmed, A.; Josifovski, V.;
  Long, J.; Shekita, E.~J.; and Su, B.-Y.
\newblock 2014.
\newblock Scaling distributed machine learning with the parameter server.
\newblock In {\em 11th $\{$USENIX$\}$ Symposium on Operating Systems Design and
  Implementation ($\{$OSDI$\}$ 14)},  583--598.

\bibitem[\protect\citeauthoryear{Lin \bgroup et al\mbox.\egroup
  }{2017}]{lin2017deep}
Lin, Y.; Han, S.; Mao, H.; Wang, Y.; and Dally, W.~J.
\newblock 2017.
\newblock Deep gradient compression: Reducing the communication bandwidth for
  distributed training.
\newblock {\em arXiv preprint arXiv:1712.01887}.

\bibitem[\protect\citeauthoryear{Mitliagkas \bgroup et al\mbox.\egroup
  }{2016}]{mitliagkas2016asynchrony}
Mitliagkas, I.; Zhang, C.; Hadjis, S.; and R{\'e}, C.
\newblock 2016.
\newblock Asynchrony begets momentum, with an application to deep learning.
\newblock In {\em 2016 54th Annual Allerton Conference on Communication,
  Control, and Computing (Allerton)},  997--1004.
\newblock IEEE.

\bibitem[\protect\citeauthoryear{Polyak and
  Juditsky}{1992}]{polyak1992acceleration}
Polyak, B.~T., and Juditsky, A.~B.
\newblock 1992.
\newblock Acceleration of stochastic approximation by averaging.
\newblock {\em SIAM Journal on Control and Optimization} 30(4):838--855.

\bibitem[\protect\citeauthoryear{Recht \bgroup et al\mbox.\egroup
  }{2011}]{recht2011hogwild}
Recht, B.; Re, C.; Wright, S.; and Niu, F.
\newblock 2011.
\newblock Hogwild: A lock-free approach to parallelizing stochastic gradient
  descent.
\newblock In {\em Advances in neural information processing systems},
  693--701.

\bibitem[\protect\citeauthoryear{Renggli \bgroup et al\mbox.\egroup
  }{2018}]{renggli2018sparcml}
Renggli, C.; Alistarh, D.; Hoefler, T.; and Aghagolzadeh, M.
\newblock 2018.
\newblock Sparcml: High-performance sparse communication for machine learning.
\newblock {\em arXiv preprint arXiv:1802.08021}.

\bibitem[\protect\citeauthoryear{Seide \bgroup et al\mbox.\egroup
  }{2014}]{seide20141}
Seide, F.; Fu, H.; Droppo, J.; Li, G.; and Yu, D.
\newblock 2014.
\newblock 1-bit stochastic gradient descent and its application to
  data-parallel distributed training of speech dnns.
\newblock In {\em Fifteenth Annual Conference of the International Speech
  Communication Association}.

\bibitem[\protect\citeauthoryear{Stich, Cordonnier, and
  Jaggi}{2018}]{stich2018sparsified}
Stich, S.~U.; Cordonnier, J.-B.; and Jaggi, M.
\newblock 2018.
\newblock Sparsified sgd with memory.
\newblock In {\em Advances in Neural Information Processing Systems},
  4447--4458.

\bibitem[\protect\citeauthoryear{Strom}{2015}]{strom2015scalable}
Strom, N.
\newblock 2015.
\newblock Scalable distributed dnn training using commodity gpu cloud
  computing.
\newblock In {\em Sixteenth Annual Conference of the International Speech
  Communication Association}.

\bibitem[\protect\citeauthoryear{Tang \bgroup et al\mbox.\egroup
  }{2019}]{tang2019doublesqueeze}
Tang, H.; Lian, X.; Zhang, T.; and Liu, J.
\newblock 2019.
\newblock Doublesqueeze: Parallel stochastic gradient descent with double-pass
  error-compensated compression.
\newblock {\em arXiv preprint arXiv:1905.05957}.

\bibitem[\protect\citeauthoryear{Tsitsiklis, Bertsekas, and
  Athans}{1986}]{tsitsiklis1986distributed}
Tsitsiklis, J.; Bertsekas, D.; and Athans, M.
\newblock 1986.
\newblock Distributed asynchronous deterministic and stochastic gradient
  optimization algorithms.
\newblock {\em IEEE transactions on automatic control} 31(9):803--812.

\bibitem[\protect\citeauthoryear{Wang and Srebro}{2017}]{wang2017stochastic}
Wang, W., and Srebro, N.
\newblock 2017.
\newblock Stochastic nonconvex optimization with large minibatches.
\newblock {\em arXiv preprint arXiv:1709.08728}.

\bibitem[\protect\citeauthoryear{Wangni \bgroup et al\mbox.\egroup
  }{2018}]{wangni2018gradient}
Wangni, J.; Wang, J.; Liu, J.; and Zhang, T.
\newblock 2018.
\newblock Gradient sparsification for communication-efficient distributed
  optimization.
\newblock In {\em Advances in Neural Information Processing Systems},
  1299--1309.

\bibitem[\protect\citeauthoryear{Wen \bgroup et al\mbox.\egroup
  }{2017}]{wen2017terngrad}
Wen, W.; Xu, C.; Yan, F.; Wu, C.; Wang, Y.; Chen, Y.; and Li, H.
\newblock 2017.
\newblock Terngrad: Ternary gradients to reduce communication in distributed
  deep learning.
\newblock In {\em Advances in neural information processing systems},
  1509--1519.

\bibitem[\protect\citeauthoryear{You, Gitman, and
  Ginsburg}{2017}]{you2017scaling}
You, Y.; Gitman, I.; and Ginsburg, B.
\newblock 2017.
\newblock Scaling sgd batch size to 32k for imagenet training.
\newblock {\em arXiv preprint arXiv:1708.03888} 6.

\bibitem[\protect\citeauthoryear{Zhang \bgroup et al\mbox.\egroup
  }{2013}]{zhang2013asynchronous}
Zhang, S.; Zhang, C.; You, Z.; Zheng, R.; and Xu, B.
\newblock 2013.
\newblock Asynchronous stochastic gradient descent for dnn training.
\newblock In {\em 2013 IEEE International Conference on Acoustics, Speech and
  Signal Processing},  6660--6663.
\newblock IEEE.

\end{thebibliography}

\end{document}